\documentstyle[epsf]{article}
\def\rp{$R_p \hspace{-1em}/\;\:$}

\def \znbb {0\nu\beta\beta}
\newcommand{\ba}[1]{\begin{eqnarray} \label{(#1)}}
\newcommand{\ea}{\end{eqnarray}}

\newcommand{\rf}[1]{(\ref{(#1)})}

\begin{document}
\begin{center}

{\bf Pion Exchange Currents in Neutrinoless Double Beta Decay\\
and Limits on Supersymmetry\footnote{
	Contribution to the Proceedings of the International Workshop
	on Non-Accelerator New Physics(NANP'97), Dubna, Russia, June 1997}
}
\medskip

{Amand Faessler$^1$,  Sergey Kovalenko$^{2}$,
Fedor \v Simkovic$^{3}$
and Joerg Schwieger$^1$
\medskip

1.{\it Institute f\"ur Theoretische Physik
         der Universit\"at T\"ubingen,\\
   Auf der Morgenstelle 14, D-72076 T\"ubingen, Germany}\\
2. {\it Joint Institute for Nuclear Research, 141980 Dubna, Russia} \\
3.{\it Department of Nuclear Physics, Comenius University, \\
   Mlynsk\'a dolina F1, 84215 Bratislava, Slovakia}
}
\end{center}

\begin{abstract}
 We examine the pion exchange mode of
neutrinoless double beta decay ($\znbb $)
induced by the $R$-parity violating quark-lepton
operators of the supersymmetric (SUSY) extensions
of the standard model of the electroweak interactions.
The corresponding nuclear matrix elements are evaluated
within the renormalized quasiparticle random phase approximation
with proton-neutron pairing, which includes
the Pauli effect of fermion pairs and does not collapse for a physical
value of the nuclear force strength.
It is argued that the pion-exchange mode of $\znbb$-decay dominates
over the  conventional two-nucleon mode in the case of the SUSY mechanism.
As a result sensitivity of $\znbb$-decay to the SUSY contribution
turns out to be  significantly better that previously expected from
the two-nucleon mode calculations.
An upper limit on the $R$-parity violating coupling
$\lambda'_{111}$ is derived from non-observation
of neutrinoless double beta decay. This limit is much stronger than
that expected from the near future accelerator experiments.
\end{abstract}

\enlargethispage{\baselineskip}

\section{Introduction}
The observation of neutrinoless nuclear double beta decay
($0\nu\beta\beta$) would undoubtedly indicate the presence of
new physics beyond the standard model (SM) of electroweak interactions.
However, as yet there is no any experimental evidence
%indication
for this lepton-number violating exotic process.
On the other hand non-observation of $\znbb$-decay at given experimental
sensitivity allows one to establish limits on some parameters of new physics.
The most famous example is the light effective Majorana neutrino mass
which is limited by the $0\nu\beta\beta$-decay experiments \cite{hdmo97}
at the level $\langle m_{M}^{\nu} \rangle \le {\cal O}(0.5-1.1 $ eV$)$
\cite{hdmo97,si97} depending on nuclear model.

It is known that the Majorana neutrino exchange is not the only possible
mechanism of $\znbb$-decay.
The lepton-number violating quark-lepton interactions
of the R-parity non-conserving supersymmetric extensions of the SM
(\rp SUSY) can also induce  this process \cite{Mohapatra,Vergados,HKK96}.
It is worthwhile to notice
that the corresponding 1st generation \rp Yukawa coupling
are so stringently constrained by non-observation of
the $0\nu\beta\beta$-process that all possible 1st generation
%the associated
\rp effects are pushed beyond the reach of the present
and the near future accelerator and the other non-accelerator experiments
\cite{HKK96,FKSS97}.

In searching for tiny effects of the physics beyond the SM
the main disadvantage of $\znbb$-decay experiments compared
to the accelerator ones is the necessity of taking into account
nuclear structure.

Although there are many difficulties the study of $\znbb$-decay
has the advantage of unprecedented accuracy and precision with
which the process can be studied  experimentally.
It is possible to observe large samples of several
kilograms of potentially decaying nuclei and to search for the decay of a
single nucleus. The lower half-life limit for $0\nu\beta\beta$-decay
measured in this way is very high. In addition, the $2\nu\beta\beta$-decay
process predicted by the standard model can be measured at the same time
in order to check the nuclear structure calculations.

The nuclear ${0\nu\beta\beta}$-decay
is triggered by the ${0\nu\beta\beta}$ quark transition
$d + d\rightarrow u + u + 2 e^-$ which is induced by certain
fundamental interactions.
It was a common practice to put the initial d-quarks separately
inside the two initial neutrons of a ${0\nu\beta\beta}$-decaying
nucleus.
This is the so called two-nucleon mode of the ${0\nu\beta\beta}$-decay.
If the above ${0\nu\beta\beta}$ quark transition proceeds at
short distances, as in the case of $R_p \hspace{-1em}/\;\:$
SUSY interactions, then the basic nucleon transition  amplitude
$n + n\rightarrow p + p + 2e^-$ is strongly suppressed
for relative distances smaller than the mean nucleon
radius.

The goal of this paper is to discuss
the pion-exchange SUSY mechanism which is
based on the one and two pion exchange between the decaying neutrons.
The two pion exchange counterpart
of this mechanism was first studied in Ref. \cite{FKSS97}.
At the quark level this mechanism implies the same short-distance
$R_p \hspace{-1em}/\;\:$
MSSM interactions as in Ref.\cite{HKK96}. However, it essentially
differs from the previous consideration of the SUSY contribution
to the ${0\nu\beta\beta}$-decay at the stage of the hadronization.
We will show that in the case of the $R_p \hspace{-1em}/\;\:$ MSSM
induced quark transition the pion-exchange contribution absolutely
dominates over the conventional two-nucleon mode.
As a result, a significant improvement of the previously known
\cite{HKK96}
two-nucleon mode $\znbb$-decay  limit  on \rp-Yukawa coupling
$\lambda'_{111}$  becomes possible.

We calculate the nuclear matrix
elements governing the two-nucleon, one pion-exchange and two-pion
exchange SUSY contributions to $\znbb$-decay within the
renormalized Quasiparticle Random Phase Approximation with
proton-neutron pairing (full-RQRPA).
This nuclear structure method has been
developed from the proton-neutron QRPA method, which has been  frequently
used in the $\znbb$-decay  calculations. The full-RQRPA is an extension
of the pn-QRPA by considering the effect of the proton-neutron pairing
and the Pauli effect of the fermion pairs in an approximate way.
In this way the sensitivity of the nuclear matrix elements to the details
of the nuclear Hamiltonian is reduced considerably and more reliable
values on the lepton number non-conserving parameters are obtained.

\section{The Nuclear Structure Method}
Most of the double beta decaying nuclei are not yet accessible for
the detailed shell model treatment
because of the complexity of this approach.
Therefore the pn-QRPA
\cite{vogel88} is mostly used for the nuclear structure
calculations.
In general the half-life for a
$0\nu\beta\beta$-process can be factorized in the form
\begin{equation}
\frac{1}{T_{1/2}}=G |ME|^2 \epsilon
\end{equation}
with the leptonic phase space factor $G$, the nuclear matrix element $ME$
and the factor $\epsilon$ that is given by the special extension of the
standard model under consideration.

The nuclear matrix element $ME$ is ruled by the nuclear structure of
the involved nuclei, but also by the considered $\znbb $-decay mechanism,
which determines  the transition operators. We calculate the matrix element
in the intermediate nucleus approach, which requires
the construction of a complete set of the intermediate nuclear
states.

In the framework of the QRPA or renormalized QRPA
the $m^{th}$ excited states with angular momentum $J$ and projection $M$
is created by a phonon-operator $Q$ with the properties
\begin{equation}
Q^{m\dagger}_{JM}|0^+_{RPA}\rangle
=|m,JM\rangle \qquad \mbox{and} \qquad
Q|0^+_{RPA}\rangle=0.
\label{stat}
\end{equation}
Here, $|0^+_{RPA}\rangle$ is the ground state of the initial or
the final nucleus.
The phonon-operator $Q$ takes the following form
\begin{equation}
Q^{m\dagger}_{JM^\pi}=\sum_{k\mu \leq l\nu}
  X^m_{(k\mu l\nu),J}A^\dagger(k\mu l\nu,JM)
+ Y^m_{(k\mu l\nu),J}\tilde{A}(k\mu l\nu,JM)
\label{phonop}
\end{equation}
where the tilde indicates time-reversal. $ A^\dagger(k\mu l\nu, JM)$
is the two quasi-particle creation and annihilation operator coupled to good
angular momentum $J$ with projection $M$ namely
\begin{equation}
A^\dagger(k\mu l\nu, JM)
= n(k\mu, l\nu) \sum^{}_{m_k , m_l }
C^{J M}_{j_k m_k j_l m_l } a^\dagger_{\mu k m_k} a^\dagger_{\nu l m_l}\,.
\end{equation}
The indices $\mu$ and $\nu$ denote the isospin structure of the
bifermion operator. In the case of a BCS definition of the
quasi-particles without pn-pairing it would distinguish between proton
and neutron type quasi-particles. For the case of a HFB definition of the
quasi-particles with pn-pairing the quasi-particles have no definite
isospin anymore and the index becomes $1$ or $2$ \cite{CBFST}. The
factor $n(k\mu, l\nu)$ guarantees the normalization for the case of
two identical particles $k\mu$ and $l\nu$.

From Eq. (\ref{stat}) and (\ref{phonop}) one can derive the RQRPA equation
\begin{equation}
\label{rpamateq}
\left(
   \begin{array}{cc}
     {\cal A} & {\cal B}\\
     {\cal B} & {\cal A}
   \end{array}
\right)_{J^\pi}
\left(
  \begin{array}{c}
    X \\
    Y
  \end{array}
\right)_{J^\pi}
=\Omega^m_{J^\pi}
\left(
   \begin{array}{cc}
     {\cal U} & 0\\
     0 & {\cal -U}
   \end{array}
\right)_{J^\pi}
\left(
  \begin{array}{c}
    X \\
    Y
  \end{array}
\right)_{J^\pi}\, ,
\end{equation}
where $\Omega^m_{J^\pi}$ denotes the excitation energy of the $m^{th}$
excited state with angular momentum $J$ and parity $\pi$ with respect
to the ground state $|0^+_{RPA}\rangle$.  The three matrices $\cal{A}$,
$\cal{B}$, $\cal{U}$ are given by the expectation values of the
following commutators in the correlated ground state
\begin{eqnarray}
\label{amat}
{\cal A}^{a\alpha b\beta}_{k\mu l\mu,J}=
\left\langle 0^+_{RPA}\right|\left[A(a\alpha b \beta,JM)
,\left[H,A^\dagger(k\mu l\nu,JM)\right]\right]\left|0^+_{RPA}\right\rangle\,,\\
\label{bmat}
{\cal B}^{a\alpha b\beta}_{k\mu l\mu,J}=
\left\langle 0^+_{RPA}\right|\left[A(a\alpha b \beta,JM)
,\left[\tilde{A}(k\mu l\nu,JM),H\right]\right]\left|0^+_{RPA}\right\rangle\,,\\
\label{umat}
{\cal U}^{a\alpha b\beta}_{k\mu l\mu,J}=
\left\langle 0^+_{RPA}\right|\left[A(a\alpha b \beta,JM)
,A^\dagger(k\mu l\nu,JM)\right]\left|0^+_{RPA}\right\rangle \,.
\end{eqnarray}
For solving the above equation it is necessary to introduce some
approximation scheme for the calculation of the  $\cal{A}$,
$\cal{B}$ and $\cal{U}$ matrices. The simplest and the most
frequently used one is the quasiboson
approximation (QBA) scheme, which implies the two-quasiparticle operator
$A^\dagger(k\mu l\nu, JM)$ to be a boson operator. However,
it is well-known that the QBA violates the Pauli exclusion
principle because we have neglected terms coming from the
commutator of the two bifermion operators by replacing the
exact expression for this commutator with the its expectation value
in the uncorrelated BCS/HFB ground state, which is the vacuum for the
quasi-particle operators.  It turns out
that the QBA  is a poor approximation and leads to too strong
ground state correlations, an unphysically large  occupation of
quasi-particle states in the correlated ground state,
close to a collapse of the QRPA solution. But the ground state
correlations influence the nuclear matrix elements severely and in
general the use of the QBA leads to a very sensitive
dependence of the nuclear matrix elements on the strength of the
residual interaction in the particle-particle channel.

To overcome this problem the Pauli-principle has to be incorporated in
the approach \cite{toiv95,SSF}, by which the occupation of the
quasi-particle states in the correlated ground state would be
limited. The Pauli-principle can be incorporated to large extend by
calculating explicitly the commutator of Eq. (\ref{umat}) with the
single quasi-particles obeying Fermi-anti-commutation rules. Neglecting
the non-diagonal part the commutator is not anymore boson like, but
obtains corrections to its bosonic behavior due to the fermionic
constituents.
\begin{eqnarray} {\cal U}^{a\alpha b\beta}_{k\mu
l\mu,J}&=&
\left\langle 0^+_{RPA}\right|\left[A(a\alpha b \beta,JM)
,A^\dagger(k\mu l\nu,JM)\right]\left|0^+_{RPA}\right\rangle \nonumber \\
&\simeq&n(k\mu, l\nu) n(a\alpha, b\beta)
\Big( \delta_{ka}\delta_{\mu \alpha}\delta_{lb}
\delta_{\nu\beta} -
\delta_{la}\delta_{\nu \alpha}\delta_{kb}
\delta_{\mu \beta}(-1)^{j_{k}+j_{l}-J}\Big) \nonumber \\
&&\times \underbrace{
\left\{
\begin{array}{ccl}
1&-&\frac{1}{\hat{\jmath}_{l}}
\langle 0^+_{RPA}|[a^\dagger_{\nu l}a_{\nu \tilde{l}}]_{00}|0^+_{RPA}\rangle \\
&-&\frac{1}{\hat{\jmath}_{k}}
\langle 0^+_{RPA}|[a^\dagger_{\mu k}a_{\mu \tilde{k}}]_{00}|0^+_{RPA}\rangle
\end{array}
\right\}
}_{
=:\displaystyle {\cal D}_{k\mu, l\nu; J^\pi}
}\,,
\label{RQBA}
\end{eqnarray}
with the known abbreviation $\hat{j}=\sqrt{(2j+1)}$. This expression is
still diagonal in the quasi-particle configuration indices, and
therefore the QRPA Eq. (\ref{rpamateq}) can easily be brought to
eigenvalue form
\begin{equation}
\label{rqrpaeq1}
  \underbrace{{\cal D}^{-1/2}\left(
    \begin{array}{cc}
       \cal A &\cal B\\
       \cal -B &\cal -A
    \end{array}
    \right)
    {\cal D}^{-1/2}}
    _{ \textstyle \overline{\cal A},\overline{\cal B}}
    \;
    \underbrace{{\cal D}^{1/2}\left(
    \begin{array}{c}
       X^m\\
       Y^m
    \end{array}
    \right)}_{\textstyle \overline{X}^m, \overline{Y}^m}
    = \Omega^m_{J^\pi}
    \underbrace{{\cal D}^{1/2}
    \left(
    \begin{array}{c}
       X^m\\
       Y^m
    \end{array}
    \right)}_{\textstyle\overline{X}^m, \overline{Y}^m}\, .
\end{equation}
As in this form the matrices $\cal{A}$ and $\cal{B}$ and the
eigenvectors are renormalized by $\cal{D}$ this approach is called
renormalized QRPA (RQRPA).  To solve this equation the quasi-particle
occupation in the correlated ground state needs to be known. But the
occupation can only be derived from the back-going amplitudes $Y$ of
the RQRPA diagonalization. An implicit formulation for the
$\cal{D}$-matrix can be given, if one express the one-body densities
of Eq. (\ref{RQBA}) in terms of mapping on two boson operators.  Then one
arrives at the formula
\begin{eqnarray}
\label{rqrpaeq2}
D_{(k\mu l\nu)J} = 1 &-& \frac{1}{\hat{\jmath}_{k^2}}\sum_{k'\mu' \atop J' m}
D_{(k\mu k'\mu')J'}\hat{J}'^2 \big |\overline{Y}^m_{(k\mu k'\mu')J'^\pi}
\big|^2
\\ \nonumber
&-&\frac{1}{\hat{\jmath}_{l^2}}\sum_{l'\nu' \atop J' m}D_{(l\nu
l'\nu')J'}\hat{J}'^2 \big |\overline{Y}^m_{(l\nu l'\nu')J'^\pi}  \big|^2 \,.
\end{eqnarray}
Eq. (\ref{rqrpaeq2}) and (\ref{rqrpaeq1}) have to be solved
self-consistently in an iterative procedure. Note that an eigenvalue
equation has to be written down separately for every multipolarity
$J^\pi$ which are then coupled by Eq. (\ref{rqrpaeq2}).
The method presented above include both proton-neutron pairing
and the Pauli effect of fermion pairs and is denoted the full-RQRPA.
In the limit the proton-neutron pairing is switched off, one obtains
the pn-RQRPA method.

In order to calculate
double beta transitions two fully independent RQRPA calculations are
needed, one to describe the beta transition from the initial to the
intermediate nucleus and another one for the beta transition from the
intermediate to the final nucleus. For that purpose the one-body transition
densities of the charge changing operator has to be evaluated. They take
the following form:
\begin{eqnarray}
<J^\pi m_i \parallel [c^+_{pk}{\tilde{c}}_{nl}]_J \parallel 0^+_i>=
\sqrt{2J+1}\sum_{\mu,\nu = 1 , 2} m(\mu k,\nu l)\times \\ \nonumber
\left [
u_{k \mu p}^{(i)} v_{l \nu n}^{(i)} {\overline{X}}^{m_i}_{\mu\nu}(k,l,J^\pi)
+v_{k \mu p}^{(i)} u_{l \nu n}^{(i)} {\overline{Y}}^{m_i}_{\mu\nu}(k,l,J^\pi)
\right ]
\sqrt{{\cal D}^{(i)}_{k \mu l \nu l J^\pi}},
\label{eq:13}
\end{eqnarray}
\begin{eqnarray}
<0_f^+ \parallel \widetilde{ [c^+_{pk'}{\tilde{c}}_{nl'}]_J}
\parallel J^\pi m_f>=
\sqrt{2J+1}\sum_{\mu,\nu=1,2}m(\mu k',\nu l')\times \\ \nonumber
\left [
v_{k' \mu p}^{(f)} u_{l' \nu n}^{(f)}
{\overline{X}}^{m_f}_{\mu\nu}(k',l',J^\pi)
+u_{k' \mu p}^{(f)} v_{l' \nu n}^{(f)}
{\overline{Y}}^{m_f}_{\mu\nu}(k',l',J^\pi)
\right ]
\sqrt{{\cal D}^{(f)}_{k' \mu l' \nu J^\pi}},
\label{eq:14}
\end{eqnarray}
with
$m(\mu a,\nu b)=\frac{1+(-1)^J\delta_{\mu \nu}\delta_{ab}}{(1+\delta_{\mu\nu}
\delta_{ab})^{1/2}}$. We note that the
${\overline{X}}^{m}_{\mu\nu}(k,l,J^\pi)$ and
${\overline{Y}}^{m}_{\mu\nu}(k,l,J^\pi)$ amplitudes are calculated
by the renormalized QRPA equation only for the configurations
$\mu a \leq \nu b$ ( i.e.,
$\mu = \nu$ and the orbitals are ordered $a \leq b$ and
$\mu = 1$, $\nu = 2$ and the orbitals are not ordered).
For different configurations
${\overline{X}}^{m}_{\mu\nu}(k,l,J^\pi)$ and
${\overline{Y}}^{m}_{\mu\nu}(k,l,J^\pi)$ in Eqs.
(\ref{eq:13}) and (\ref{eq:14}) are given by following the prescription
in Eqs. (65) and (66) of Ref. \cite{pa96}.
The index i (f) indicates that the quasiparticles and the excited
states of the nucleus are defined with respect to the initial (final)
nuclear ground state $|0^+_i>$ ($|0^+_f>$). $c^+_p$ and $ c_n$ denote
proton particle creation and neutron hole annihilation operators,
respectively.

For a complete presentation of the renormalized QRPA  method
we have to discuss also limitations of this nuclear structure method.
There are no doubts that
RQRPA offers advantages over QRPA:\\
i) There is no collapse of RQRPA
solution for a physical value of the particle-particle interaction
strength.\\
ii) The RQRPA Hamiltonians demonstrate better mutual
correspondence like the QRPA ones \cite{du97}.
Nevertheless the RQRPA has also several shortcomings:\\
i) In the framework of the RQRPA the Ikeda sum rule \cite{ikeda63}
(here given for Gamow-Teller $\beta$-transitions)
\begin{eqnarray}
S=S_-(1^+)-S_+(1^+)&=&
\frac{1}{2J+1} \sum_m |\langle m, 1^+|\beta_-|0\rangle|^2 \\ \nonumber
&-&\frac{1}{2J+1}
\sum_m|\langle m, 1^+|\beta_+|0\rangle|^2 \\ [2ex]
&=&3(N-Z)\,
\end{eqnarray}
is not fulfilled. It is worthwhile to notice that the
completeness of the excited $1^+$ states is the only
condition needed to derive the Ikeda sum rule.
In the QRPA this sum rule is identically
fulfilled, while in RQRPA the violation
is proportional to the renormalizing matrix $\cal{D}$, which
depends  on the strength of the residual interaction.
In the framework of the RQRPA the Ikeda sum rule is exhausted only
to  70-80\%. It is supposed that the omission of the scattering terms
in both the construction of the excited states and the evaluation of
one-body densities could be the reason of it.
At the moment there is work under progress
including these operators in the framework of extended RQRPA
\cite{raduta97}. There is a hope that in such a model
the Ikeda sum rule could be restored.\\
ii) The two sets of intermediate nuclear states generated from the
initial and final ground states are not identical because
of the considered QBA or RQBA schemes. Therefore the overlap factor
of these states is introduced in the theory as follows\cite{du97}:
\begin{eqnarray}
<J_{m^{}_{f}}^+ | J_{m^{}_{i}}^+> \approx
[Q^{ m_f}_{JM}, Q^{+ m_i}_{JM}] \approx
\sum_{\mu k \leq \nu l,~~ \mu' k' \leq \nu' l'}
\delta_{kk'}\delta_{ll'}{\tilde{u}}_{k\mu\mu'} {\tilde{u}}_{l\nu\nu'}
\times \\ \nonumber
\big(
\overline{X}^{m_{i}^{}}_{\mu \nu}(k, l, J)
\overline{X}^{m_{f}^{}}_{\mu' \nu'}(k, l, J)-
\overline{Y}^{m_{i}^{}}_{\mu \nu}(k, l, J)
\overline{Y}^{m_{f}^{}}_{\mu' \nu'}(k, l, J)
\big),
\end{eqnarray}
with
\begin{equation}
{\tilde{u}}_{k \mu \mu'} = u^{(i)}_{k\mu}u^{(f)}_{k\mu'} +
v^{(i)}_{k\mu}v^{(f)}_{k\mu'}.
\end{equation}
Here, $Q^{ m_f}_{JM}$  and $Q^{+ m_i}_{JM}$ are respectively
phonon annihilation and creation operators for the initial
and final nuclear states. We note that in the previous calculations
$\tilde{u}$ was approximated by the unity. However, in that case
the overlap factor is dependent on the phases of the occupation
BCS/HFB amplitudes u's and v's, which are in principal arbitrary.
A negligible difference between the results with the above overlap
containing $\tilde{u}$ and  with the overlap without $\tilde{y}$
one obtains only if the phases of the BCS/HFB amplitudes are
chosen so that $\tilde{u}$ is positive for each level.

\section{\rp SUSY induced $\znbb$-decay.
         Matrix elements and experimental constraints.}
\begin{figure}
\begin{center}
\leavevmode
\epsfxsize=4cm
\epsfbox{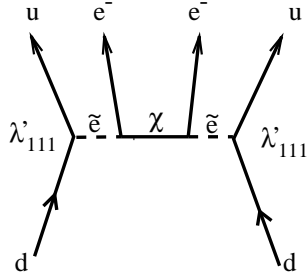}
\end{center}
\caption{\label{susycont}%
One example diagram for a possible contribution to neutrinoless double
beta decay in $R$-parity breaking
SUSY-extensions of the standard model.}
\end{figure}
$R$-parity is a discrete multiplicative symmetry
defined as $R_p=(-1)^{3B+L+2S}$,
where $S,\ B$ and $L$
are the spin, the baryon and the lepton quantum number.
This symmetry is
conserved in the minimal supersymmetric models (MSSM).
A consequence
of this symmetry would be, that SUSY-partners can only be produced in
associated production and the lightest SUSY particle is stable.

However, $R$-parity might not be conserved.
We consider presently popular  models with the explicit $R_p$-violation.
The $R_p$-violating part of the superpotential
breaking lepton number conservation and relevant to $\znbb$-decay
is
\ba{R-viol}
W_{R_p \hspace{-0.8em}/\;\:} = \lambda'_{ijk}L_i Q_j {\bar D}_k.
\ea
Here $L$,  $Q$ are lepton and quark
doublets while  ${\bar E}, \ {\bar U},\  {\bar D}$
are lepton and {\em up}, {\em down} quark singlet superfields.
Indices $i,j, k$ denote generations.
An example of the diagram contributing to $\znbb$-decay
is shown in Fig. \ref{susycont}. It involves the lepton number
violating interactions originated from the superpotential
\rf{R-viol}.

\begin{figure}
\begin{center}
\leavevmode
\epsfxsize=\textwidth
\epsfbox{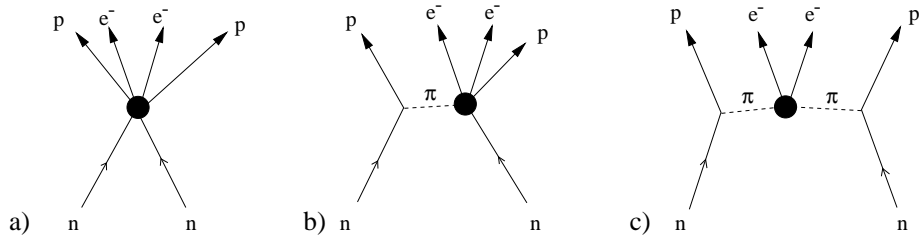}
\end{center}
\caption{\label{su_pro_cont}%
Two-nucleon (a), one-pion (b) and two-pion (c) mode of neutrinoless double
beta decay in $R$-parity violating SUSY extensions of the standard model.
%The interaction vertex is a complicated effective vertex.
}
\end{figure}

The supersymmetric model only gives the underlying transition of a
down-quark to an up-quark. Then this transition has to be transformed to
one going from a neutron to a proton. Looks natural and most
straightforward to incorporate the quarks in the nucleons.
In this way, we come up with the well known two-nucleon mode.
The corresponding effective \rp SUSY induced
nucleon-nucleon interaction is shown in Fig.
\ref{su_pro_cont}a).
But the intermediate SUSY-partners are very heavy particles, and
therefore in the two-nucleon mode the two decaying neutrons must come
very close to each other, what is suppressed by the nucleon repulsion.

Another possibility is to incorporate quarks undergoing the
\rp SUSY transition $d + d\longrightarrow u + u + 2e^-$  not
in nucleons but in virtual pions \cite{FKSS97}.
%As there are two independent quarks in
%the process, there are also two possibilities to set up pion
%contributions.
There are two possibilities to set up the pion contributions.
Namely,  only one quark-antiquark  pair $\bar{u}d$
is placed in an intermediate pion leading to a diagram shown
in Fig. \ref{su_pro_cont}b) or both $\bar{u}d$ pairs are placed
separately in two intermediate pions, as in  Fig. \ref{su_pro_cont}c).
The scale on which the
pion-contribution is enhanced compared to the two-nucleon mode is the
ratio of the nuclear form factor cut-off to the pion mass. As this
factor favors for each nuclear decay process the pion emission, the
two-pion mode is expected to be dominant over the two-nucleon and also the
one-pion mode.

The half-life for neutrinoless double beta decay regarding all the three
possibilities of hadronization of the quarks can be written in the
form
\begin{eqnarray}
\label{susyhalflife}
\lefteqn{\big[ T_{1/2}^{0\nu}(0^+ \rightarrow 0^+) \big]^{-1}
}\\ \nonumber
&=&
G_{01} \left(\frac{m_A}{m_{_p}}\right)^4
\left | \eta_{\tilde q}\cdot {\cal M}_{\tilde q}^{2N}
 +  \eta_{\tilde f}\cdot {\cal M}_{\tilde f}^{2N} +
(\eta_{\tilde q} + \eta_{\tilde f})\cdot {\cal M}^{\pi N} \right |^2\,.
\end{eqnarray}
Here $G_{01}$ is the standard phase space factor tabulated
for various nuclei in Ref. \cite{pa96} and $m_A = 850$ MeV.
The factors $\eta_{\tilde q}$ and $\eta_{\tilde f}$ are ruled by  the
SUSY-parameters and by the strength of $R$-parity violating coupling.
This factors will be constrained by the non-observation of
neutrinoless double beta decay, and under reasonable assumptions for
the SUSY-parameters (especially the SUSY masses), a limit on the
$R$-parity violating coupling $\lambda'_{111}$ can be derived.
The nuclear matrix elements ${\cal M}_{\tilde q,\tilde f }^{2N}$
governing the two-nucleon mode were presented  in Ref.\cite{HKK96}.
The one- and two-pion modes contribute to the nuclear matrix element
${\cal M}^{\pi N}$ which consists of the four partial matrix elements
\begin{equation}
\label{susy2pime}
{\cal M}^{\pi N} = \frac{m_{_p}}{ m_e}\Big[
      \alpha^{1\pi}\left(M_{GT,1\pi} + M_{T,1\pi} \right)
      +
      \alpha^{2\pi}\left(M_{GT,2\pi} + M_{T,2\pi} \right)\Big]\,,
\end{equation}
two (Gamow-Teller and Tensor type) elements for each pionic mode.
The two types of matrix elements are given by the expression
\begin{eqnarray}
M_{GT,k\pi} &=&
\langle 0^+_f|\sum_{i\neq j} \tau_i^+ \tau_j^+
\vec{\sigma}_i \cdot \vec{\sigma}_j
\frac{F_1^{(k)}(x_{\pi})R}{|\vec{r}_i-\vec{r}_j|}
|0^+_i\rangle\,,\quad\mbox{with}\quad k=1,2\\ \nonumber
M_{T,k\pi}&=&
\langle 0^+_f|\sum_{i\neq j} \tau_i^+ \tau_j^+
\left[3(\vec{\sigma}_i\cdot \hat{\vec{r}}_{ij})
       (\vec{\sigma_j}\cdot \hat{\vec{r}}_{ij})
      - \vec{\sigma}_i\cdot \vec{\sigma_j}\right]
      \frac{F_{2}^{(k)}(x_{\pi}) R}{|\vec{r}_i-\vec{r}_j|}
|0^+_i\rangle \,,
\end{eqnarray}
with $x_\pi=m_\pi r_{ij}$, $r_{ij}=|\vec{r}_i-\vec{r}_j|$ and
$\hat{\vec{r}}_{ij}=(\vec{r}_i-\vec{r}_j)/r_{ij}$.  The structure
functions $F_1^{(k)}(x_{\pi})$ and $F_2^{(k)}(x_{\pi})$ have their
origin in the integration of the propagator of the intermediate
particles and take the following form:
\begin{eqnarray}
F_1^{(1)}(x) &=&  e^{- x}, \ \ \ F_2^{(1)}(x) =
(3 + 3x + x^2)\frac{e^{- x}}{x^2},\\
F_1^{(2)}(x) &=& (x - 2) e^{- x}, \ \ \ F_2^{(2)}(x) = (x + 1) e^{- x}.
\label{potentials}
\end{eqnarray}
The structure coefficients $\alpha^{1\pi}$ and
$\alpha^{2\pi}$ depend on the hadronization of the quarks. For the
two-pion coefficient $\alpha^{2\pi}$ a hadronic matrix element of the
type $\langle\pi^+|J_iJ_i|\pi^-\rangle$ ($i=P,S,T$) for the hadronic
currents $J_i$ needs to be evaluated. This can be done in either a
non-relativistic quark model (QM) or by inserting the vacuum state in
between the two currents (vacuum insertion approximation, VIA). As the
VIA neglects contributions from other intermediate states than the
vacuum, it will give a more conservative limit. The values for the
structure coefficients $\alpha^{1\pi}=-4.4\cdot10^{-2}$ and
$\alpha^{2\pi}=0.2\mbox{(VIA), }0.64\mbox{(QM)}$ reveal the above made
statement on the dominance of  the two-pion over the one-pion mode.
Further it will be seen that not only the structure coefficient
but also the corresponding nuclear matrix elements
$M_{GT,k\pi}$ and $M_{T,k\pi}$ of the two-pion mode is larger
than those of the one-pion mode.

For further calculation  the nuclear matrix element governing the
SUSY-mechanism of $0\nu\beta\beta$-decay are transformed to ones
containing two-body matrix element in relative coordinate. One arrives
at the expression:
\begin{eqnarray}
<O_{12}>=
\sum_{{k l \acute{k} \acute{l} } \atop {J^{\pi}
m_i m_f {\cal J}  }}
~(-)^{j_{l}+j_{k'}+J+{\cal J}}(2{\cal J}+1)
\left\{
\begin{array}{ccc}
j_k &j_l &J\\
j_{l'}&j_{k'}&{\cal J}
\end{array}
\right\}\times ~~~~~\nonumber \\
< 0_f^+ \parallel
\widetilde{[c^+_{pk'}{\tilde{c}}_{nl'}]_J} \parallel J^\pi m_f>
<J^\pi m_f|J^\pi m_i>
<J^\pi m_i \parallel [c^+_{pk}{\tilde{c}}_{nl}]_J \parallel
0^+_i >\nonumber \\
\times<pk,pk';{\cal J}|f(r_{12})\tau_1^+ \tau_2^+ {\cal O}_{12}
f(r_{12})|nl,nl';{\cal J}>.~~~~~~~~~~
\end{eqnarray}
with a short-range correlation function
\begin{equation}
\label{2nuccorr}
f(r)=1-e^{-\alpha r^2 }(1-b r^2) \quad \mbox{with} \quad
\alpha=1.1 \mbox{fm}^2 \quad \mbox{and} \quad  b=0.68 \mbox{fm}^2,
\end{equation}
which takes into account the short range repulsion of the nucleons.

\begin{table}[t]
\caption{Nuclear matrix elements for SUSY two-nucleon, one-pion
and two-pion mechanisms of neutrinoless double beta decay for
 $^{76}Ge(0^{+}) \rightarrow ^{76}Se({0^{+}})$ nuclear transition
within the pn-RQRPA and full-RQRPA. The
12-level (the full $2-4\hbar\omega$ major oscillator shells)
model space has been considered. The presented values has been
calculated for $g_{pp}=1.0$. The BM, QM and VIA  denote bag model,
non-relativistic quark model and vacuum  insertion approximation,
respectively.}
\label{table1}
\begin{tabular}{cccccccc}\hline
 & & & & & & & \\
mechan. & \multicolumn{6}{c}{ two-nucleon}\\
 & & & & & & & \\ \cline{2-8}
M.E. & $M_{GT,N}$ & $M_{F,N}$ & $M_{GT'}$ & $M_{F'}$ & $M_{T'}$
& $M_{\tilde{q}}$ & $M_{\tilde{f}}$ \\
 & [$10^{-2}$] & [$10^{-2}$] & [$10^{-2}$] & [$10^{-3}$]
& [$10^{-3}$] & (BM) & (BM) \\
 & & & & & &(NQM) & (NQM) \\ \hline
pn-   & 7.05 & -2.48 & -1.04 & 3.76 & -2.38  & -116  & 5.7 \\
RQRPA & & & & & & -156 & 2.7 \\
full- & 4.32 & -1.26 & -0.65 & 1.94 & -0.76 & -75 & 5.5 \\
RQRPA & & & & & &-95 & 4.2 \\
 & & & & & & &  \\ \hline
 & & & & & & &  \\
mechan. & \multicolumn{3}{c}{ one-pion} & \multicolumn{3}{c}{ two-pion} & pion
\\ \cline{2-8}
M.E. & $M_{GT,1\pi}$ & $M_{F,1\pi}$ & $M_{1\pi}$ & $M_{GT,2\pi}$ &
$M_{F,2\pi}$
& $M_{2\pi}$ & $M^{\pi N}$ \\
 & & & & & & (VIA) & (VIA) \\
 & & & & & & (QM) & (QM) \\
 & & & & & & & \\ \hline
pn-   & 1.296 & -1.023 & -22 & -1.341 & -0.653 & -1.99 & -754\\
RQRPA & & & & & & & -2364 \\
full- & 0.840 & -0.450 & -31 & -0.810 & -0.385 & -439 & -470 \\
RQRPA & & & & & & -1403 & -1434 \\
 & & & & & & &  \\ \hline
\end{tabular}
\end{table}

The calculated nuclear matrix elements for the $0\nu\beta\beta$-decay
of A=76 isotope within the pn-RQRPA and the
full-RQRPA are presented in Table 1. The considered single-particle
model space has been  the 12-level model space
(the full $2-4\hbar\omega$ major oscillator shells)
introduced in Ref.\cite{si97}.
One should note that the calculation of the matrix
elements in QRPA collapses with increasing interaction strength, while
a calculation with RQRPA stays stable in the whole range of physical
interaction strength. The  nuclear matrix elements listed in Table 1 have
been obtained for the $g_{pp}=1$ ($g_{pp}$- parameter used to
renormalized particle-particle interaction of the nuclear Hamiltonian).
By glancing the Table 1 we note less important role of the one-pion
exchange SUSY mechanism and that  the two-pion exchange nuclear
matrix elements clearly dominate over nuclear matrix elements of
 the two-nucleon mechanism.
There are two sources of this enhancement. The first source has pure
nuclear origin
as the potential determined by the pion-exchange with mass about
140 MeV is favored in
comparison with the potential determined by the cut-off with value
about 850 MeV. The second source of enhancement has its origin
in the hadronization of the \rp SUSY effective vertex operator
$\bar{u}\gamma_5 d\cdot\bar{u}\gamma_5 d\cdot \bar{e}P_R e^c$ replaced by
its hadronic image $\pi^2\cdot  \bar{e}P_R e^c$.
The enhancement occurs due to the coincidence
of the pseudoscalar quark bilinears $\bar{u}\gamma_5 d$ with
$\pi$-meson field.

For deducing constraints
on the $R$-parity violating SUSY parameters  from
the non-observation of $\znbb$-decay we shall
use values of nuclear matrix elements
obtained within the full-RQRPA.
The current experimental lower bound on the $^{76}$Ge
${0\nu\beta\beta}$-decay half-life \cite{ba97}  is
\ba{exp}
T_{1/2}^{{0\nu\beta\beta}-\mbox{exp}}(0^+ \rightarrow 0^+)
\hskip2mm \geq \hskip2mm
1.1 \times 10^{25} \mbox{ years} \ \ \ \ \ \ \ \ \ 90 \% \ \mbox{c.l.}
\ea
With the above calculated nuclear matrix elements lower
limit can be transformed to the following upper limits for
the 1st generation \rp Yukawa coupling constant
\ba{limits}
\lambda'_{111} &\leq & 1.3 (0.8) 10^{-4}\Big({m_{\tilde q}\over{100 GeV}}
\Big)^2
 \Big({m_{\tilde g}\over{100 GeV}} \Big)^{1/2}\\
\label{L2}
\lambda'_{111}&\leq& 9.1 (5.2) 10^{-4}
\Big({m_{\tilde e}\over{100 GeV}} \Big)^2
 \Big({m_{\chi}\over{100 GeV}} \Big)^{1/2}
\ea
We point out at that the uncertainties of
the nuclear structure calculations are smaller
than those from the hadronic matrix elements.
The limits in Eqs. \rf{limits}-(\ref{L2}) correspond to
the VIA (QM) calculations of the hadronic matrix element.

These limits are much stronger than those previously known
and lie beyond the reach of the near future  accelerator experiments
(though, accelerator experiments are potentially sensitive
to other couplings than $\lambda'_{111}$).  To constrain the
size of $\lambda'_{111}$ one needs to make assumptions on the masses
of the SUSY-partners. If the masses of the SUSY-partners would be at
their present limit \cite{RPP}, one could constrain the coupling to
$\lambda'_{111}\leq 6.0 (3.3) \cdot 10^{-5}$. A conservative bound can
be set by assuming all the SUSY-masses being at the "SUSY-naturalness"
bound of 1~TeV, leading to $\lambda'_{111}\leq 8.2 (4.4)\cdot
10^{-2}$.

These results show that the non-observation of $0\nu\beta\beta$-decay
strongly limits extensions of the standard model of electroweak
interaction. Although a many-body problem needs to be solved the
improvement of the limits is so large that it overcomes the
uncertainties
in the nuclear and hadronic
matrix elements and leads to limits that are much stronger than those
from accelerator and non-accelerator experiments.

%\section*{References}

\end{document}